\documentclass[a4paper]{svjour3}
\usepackage{graphicx}
\usepackage{amsmath}
\usepackage{cite}%
\begin{document}

\title{Variable Chaplygin gas cosmologies in $f(R,T)$ gravity with particle creation}
%\subtitle{Do you have a subtitle?\\ If so, write it here}

%\titlerunning{Short form of title}        % if too long for running head

\author{N. Hulke \and G. P. Singh \and Binaya K. Bishi \and A. Singh  }

%\authorrunning{Short form of author list} % if too long for running head

\institute{N. Hulke \& G. P. Singh \at
              Department of Mathematics,  Visvesvaraya National Institute of Technology, Nagpur, India\\
              \email{nik19hulke@gmail.com,gpsingh@mth.vnit.ac.in}           %  \\
%             \emph{Present address:} of F. Author  %  if needed
           \and
           Binaya K. Bishi  \at
           Department of Mathematics, Lovely Professional University, Jalandhar, India \\
           \email{binaybc@gmail.com}
           \and
          Ashutosh Singh \at
          Department of Applied Mathematics, Jabalpur Engineering College, Jabalpur, India \\
           \email{theprabhu.09@gmail.com}
           \and
          N. Hulke and Binaya K. Bishi contributed equally to this work.Authors to whom correspondence should be addressed. Electronic addresses: nik19hulke@gmail.com and binaybc@gmail.com
}

\date{Received: date / Accepted: date}
% The correct dates will be entered by the editor

\maketitle

\begin{abstract}
A flat FLRW cosmological model with perfect fluid comprising of variable Chaplygin gas has been studied in context of $f(R,T)$ gravity with particle creation. The considered scale factors describe the accelerated expansion of universe due to the effective negative pressure produced during evolution of universe. The role of particle creation pressure on the cosmological parameters have been discussed in detail. By considering well accepted values of free parameters, the late time expansion of the universe with energy conditions have also been studied. The state-finder diagnostic for the considered cases have been studied and the evolution of source function with time have suggested almost constant particle production at late times.
\keywords{Intermediate law \and emergent law \and accelerated universe \and  particle creation }
% \PACS{PACS code1 \and PACS code2 \and more}
% \subclass{MSC code1 \and MSC code2 \and more}
\end{abstract}

\section{Introduction}
$f(R)$ gravity introduces coupling-free functions which helps to resolve many cosmological issues like the accommodation of recent observational predictions \cite{1,2,3,4}. The idea of a non-minimal curvature matter coupling successfully incorporates cluster of galaxies, yielding natural preheating conditions corresponding to inflationary models \cite{5,6,7,8}. A new version of modified gravity known as $f(R,T)$ gravity incorporates curvature as well as matter induces strong interactions of gravity and matter, and thus it can used for the explanation of current accelerating scenario of the universe \cite{9}. The cosmological investigations of issues like expansion scalar, deceleration parameter, energy conditions, thermodynamics, exact solutions having anisotropic nature, reconstruction and stability of dark energy models has been done in literature \cite{10,11,12,13,14,15,16,17,18,19,20,21,22,23}. \\
The gravitational particle production theory has a long history. In the literature, special attention has been attracted by adiabatic production of perfect fluid particles where the entropy per particle (specific entropy) is conserved \cite{24,25,26,27,28}. Due to the enlargement of phase space of system, there is overall particle production as the particle number increases. The condition of conservation of specific entropy during the production of fluid particles leads to a relation between particle creation pressure and particle production rate. The idea of particle production in different theories of gravity have been extensively studied in literature \cite{26,27,28,29,30,31,32,33,34,35,36,37,38,39,40,41,42,43,44,45,46,47,48,49}. \\
The barotropic fluids and extensions of barotropic equation of state $p=\gamma\rho$ which are built up with different equation of state \cite{50,51,52,53,54,55,56,57,58,59,60,61,62,63,64} presents relevant examples of perfect fluids. The forms of barotropic fluids dubbed as Chaplygin gas \cite{53,54} and generalizations of Chaplygin gas \cite{50,51,52} have also been used in literature to interpolate between an early time decelerated phase and late time accelerated expansion phase. The cosmological features of variable Chaplygin gas described a unified dark energy - matter scenario have been studied in five dimensional Kaluza-Klein gravity \cite{65}.
\par In this paper, we are studying particle creation mechanism in the framework of $f(R,T)$ gravity with variable Chaplygin gas in the flat FLRW geometrical background. In section 2, we write cosmological equations of $f(R,T)$ gravity with particle production and obtain the cosmological solutions by assuming the antasz  of scale factor given by $a(t)=\frac{-1}{t}+t^{2}$, $a(t)=a_{0}(k+e^{\mu t})^{\nu}$ and $a(t)=exp(mt^{l})$. In section 3, we discuss the general characteristics and issues of the proposed models with the observational characteristics of the universe. In section 4, we conclude the results and our understanding of the given models.
\section{$f(R,T)=R+2f_1(T)$ gravity with particle creation and the cosmological solutions}
$f(R,T)$ theory of gravity is the modification of General Relativity (GR). The equations of motion in the $f (R, T)$ theories of gravity, where $R$ and $T$ are scalar curvature and the trace of energy-momentum tensor respectively \cite{9}, can be described by total gravitational action of the form
\begin{equation}
S=\frac{1}{16\pi G}\int\sqrt{-g}\left[\ \textit{f(R,T)}+L_{m}\right]d^{4}x
\label{eq1}
\end{equation}
where $g$ is the determinant of metric tensor,  $L_{m}$ be the matter Lagrangian density. Varying the action (\ref{eq1}) with respect to metric tensor yields
\begin{equation}
G_{\mu\nu}=[8\pi+2\ \textit{f}\ ^{'}(T)]\ T_{\mu\nu}+2\ [\textit{f}\ ^{'}(T)p+\frac{1}{2}\ \textit{f}(T)]g_{\mu\nu},
\label{eq2}
\end{equation}
In this paper, we assume $f(R,T)=R+2f_1(T)$. By prime, we denote differentiation with respect to the argument. We take the simplest non-trivial functional form of the function in ${f}(R,T)$ gravity given by $f_1(T)=\lambda T$, where $\lambda$ is a constant \cite{9}. The energy momentum tensor of perfect fluid can be written as,
\begin{equation}
T_{\mu\nu}=(\rho+p_{m})u_{\mu}u_{\nu}-p_{m}g_{\mu\nu}
\label{eq3}
\end{equation}
where $\rho$ and $p_{m}$ defines the energy density and pressure of the cosmic fluid respectively. $u^{\mu}$ represents the components of the four velocity vector in the co-moving co-ordinate system which satisfies the condition $u^{\mu}u_{\mu}=1$. We choose the perfect fluid matter as $L_{m}=-p_{m}$ in the action (\ref{eq1}). And, we next consider the matter source of the universe to be perfect fluid with equation of state of variable Chaplygin gas \cite{50,51,52}. \\
We take the flat FLRW metric given by,
\begin{equation}
ds^{2}=dt^{2}-a^{2}(t)\sum_{i=1}^{3}(dx_{i}^{2})
\label{eq4}
\end{equation}
where, $a(t)$ is the scale factor. The energy momentum tensor of the perfect fluid (\ref{eq3}) in the presence of particle creation takes the form \cite{32,33,34}
\begin{equation}
T_{\mu\nu}=(\rho+p_{m}+p_{c})u_{\mu}u_{\nu}-(p_{m}+p_{c})g_{\mu\nu}
\label{eq5}
\end{equation}
where $p_{c}$ is the particle creation pressure which depends on the particle production rate. The trace of the energy momentum tensor (\ref{eq5}) is given as
\begin{equation}
T=\rho-3(p_{m}+p_{c})
\label{eq6}
\end{equation}
The gravitational field equations in the flat FRW background can be written as,
\begin{equation}
3H^{2}=8\pi\rho+\textit{f}(T)+2(\rho+p_{m}+p_{c})\ \textit{f\ }^{'}(T)
\label{eq7}
\end{equation}
\begin{equation}
2\dot{H}+3H^{2}=-8\pi(p_{m}+p_{c})+\textit{f}(T)
\label{eq8}
\end{equation}
where an overhead dot denotes the derivative with respect to cosmic time $t$. Hubble parameter is defined as $H=\frac{\dot{a}}{a}$.
Equations (\ref{eq6}), (\ref{eq7}) and (\ref{eq8}) yields
\begin{equation}
3H^{2}=(8\pi+3\lambda)\rho-\lambda p_{m}-\lambda p_{c}
\label{eq9}
\end{equation}
\begin{equation}
2\dot{H}+3H^{2}=\lambda\rho-(8\pi+3\lambda)p_{m}-(8\pi+3\lambda)p_{c}
\label{eq10}
\end{equation}
The accelerating expansion of the universe is compiled by the negative $p_{c}$. The radiation component has no influence on the acceleration and the ordinary particle production is much limited because of the firm constraints foist by local gravity measurements \cite{29,30,31}.
\par Adiabatic particle production means particle as well as the entropy $(S)$ (with entropy per particle $(\sigma=\frac{S}{N})$ being constant) have been produced in the space-time. The creation pressure, in case of conserved specific entropy $\sigma$ (that is, entropy per particle $(\sigma=\frac{S}{N})$), is given by \cite{26,27,35}
\begin{equation}
p_{c}=-\frac{(\rho+p_{m})\Gamma}{3nH}
\label{eq11}
\end{equation}
The entropy $S$ is not conserved due to the enlargement of phase space resulting from the particle production \cite{28}. Here we use the parametrization of source function $\Gamma$ \cite{36,37,38,39,40,41,42,43,47} as $\Gamma=3nH\eta$. If $\Gamma>0$, then there is particle production; $\Gamma<0$ indicates particle annihilation and vanishing $\Gamma$ shows there is no particle production. Particle number density is denoted by $n$ and $0\leq\eta\leq 1$ is a constant. The term $n\eta>0$ characterises the particle production process and can act as a free parameter in the model. $n\eta=0$ indicates no particle creation whereas high $n\eta$ shows high particle production rate. For all these above cases, $\frac{\Gamma}{3H}\leq 1$. Using $\Gamma=3nH\eta$, the particle creation pressure $p_{c}$ reduces to
\begin{equation}
p_{c}=-(\rho+p_{m})\eta
\label{eq12}
\end{equation}
By using equation (\ref{eq12}) in equations (\ref{eq9}) and (\ref{eq10}), we get
\begin{equation}
3H^{2}=[8\pi+(3+\eta)\lambda]\rho+\lambda(\eta-1)p_{m}
\label{eq13}
\end{equation}
\begin{equation}
2\dot{H}+3H^{2}=[8\pi\eta+(1+3\eta)\lambda]\rho+[8\pi(\eta-1)+3\lambda(\eta-1)]p_{m}
\label{eq14}
\end{equation}
Above field equations (\ref{eq13}) and (\ref{eq14}) contains $a$ in the Hubble parameter and its derivative. So, we have two field equations containing three variables $\rho$, $p_{m}$ and $a$. Now, we consider the variable Chaplygin gas \cite{50,51,52} with equation of state
\begin{equation}
p_{m}=\frac{-Ba^{-n}}{\rho}
\label{eq15}
\end{equation}
where $B$ is any positive constant, $\rho$ is the energy density and $a$ is the scale factor. The variable Chaplygin gas is a phenomenologically extended model of Chaplygin gas. This form of energy density interpolates between an early time decelerated expansion phase and a late times accelerated expansion phase. Using EoS (\ref{eq15}) of variable Chaplygin gas (VCG), the field equations (\ref{eq13}) and (\ref{eq14}) reduces into
\begin{equation}
3H^{2}=[8\pi+(3+\eta)\lambda]\rho+B\lambda(\eta-1)a^{-n}\rho^{-1}
\label{eq16}
\end{equation}
\begin{equation}
2\dot{H}+3H^{2}=[8\pi\eta+(1+3\eta)\lambda]\rho-[8\pi(\eta-1)+3\lambda(\eta-1)]Ba^{-n}\rho^{-1}
\label{eq17}
\end{equation}
Eliminating $H^{2}$ from equations (\ref{eq16}) and (\ref{eq17}), a single evolution equation can be written as,
\begin{equation}
\dot{H}=(\eta-1)(4\pi+\lambda)\rho-(\eta-1)(4\pi+\lambda)Ba^{-n}\rho^{-1}
\label{eq18}
\end{equation}
Alternatively, one can write equation (\ref{eq18}) as a polynomial equation in term of $\rho$ given by
\begin{equation}
\rho^{2}+\left[\frac{\dot{H}}{(\eta-1)(4\pi+\lambda)}\right]\rho-Ba^{-n}=0
\label{eq19}
\end{equation}
The recent observational data show that Universe is expanding with acceleration. In next few subsections, we discuss three different physically viable cosmologies in which the scale factors are describing the accelerated phases of the Universe.
\subsection{Model-I}
\begin{figure}[ht!]
\centering
\includegraphics[width=15cm]{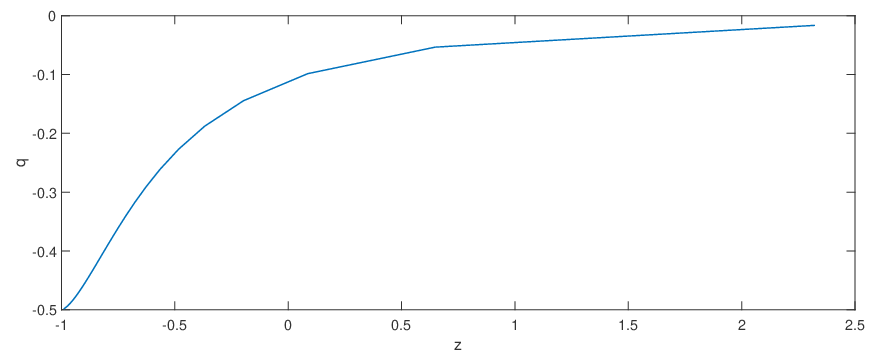}
\caption{Variation of $q$ versus $z$}\label{fig1}
\end{figure}

We consider the scale factor $a(t)$ of form
\begin{equation}
a=\frac{-1}{t}+t^{2},
\label{eq20}
\end{equation}
The choice of above form of scale factor yields a time dependent deceleration parameter. It is interesting to note that as $t\rightarrow 0$, $H\rightarrow 0$ and therefore the inflationary scenario at early stages of the universe can be observed. A. Pradhan \cite{66} have studied the above \textit{ansatz} (\ref{eq20}) in Bianchi type-$VI_{o}$ space-time in the study of the accelerating dark energy models with anisotropic fluid. Pradhan et.al. \cite{67} have studied the Bianchi-I cosmological models in scalar-tensor theory of gravitation with the above form of scale factor (\ref{eq20}). Amirhashchi et.al. \cite{68} used the relation (\ref{eq20}) in investigation of the FRW universe with interacting and non interacting two-fluid atmosphere for dark energy. Kotmbkar et.al. \cite{69} have considered the expression (\ref{eq20}) in Bianchi type-I space-time in the presence of generalized Chaplygin gas to investigate the evolution of universe with cosmological constant $\Lambda$ and gravitational constant $G$. The Hubble parameter takes the form
\begin{equation}
H=\frac{2t^{3}+1}{t(t^{3}-1)}
\label{eq21}
\end{equation}
Deceleration parameter for scale factor (\ref{eq20}) is given by
\begin{equation}
q=-2\frac{(t^{3}-1)^{2}}{(1+2t^{3})^{2}}
\label{eq22}
\end{equation}
The negative values of $q$ gives information about the general dynamical behaviour of the universe: deSitter expansion happens at $q =-1$, accelerating power-law expansion can be achieved for $-1 < q \leq 0$ and a super-exponential expansion happens for $q <-1$. In the present model, throughout in it's evolution, universe have $-1\leq q \leq 0$ and at late times $q\rightarrow -0.5$ and thus satisfies the condition for the accelerating universe. The behaviour of $q$ versus $z$ has been plotted in figure \ref{fig1}.

Equation (\ref{eq19}) is a quadratic equation in $\rho$.  We get the roots of $\rho$ in terms of $t$ as
\begin{equation}
\rho=-\left[\frac{(1-2t^{6}-8t^{3})}{2t^{2}(t^{3}-1)^{2}(\eta-1)(4\pi+\lambda)}\right]\pm\left[\frac{(1-2t^{6}-8t^{3})^{2}}{4t^{4}(t^{3}-1)^{4}(\eta-1)^{2}(4\pi+\lambda)^{2}}+B\frac{t^{n}}{(t^{3}-1)^{n}}\right]^\frac{1}{2}
\label{eq23}
\end{equation}
$\rho$ can not be negative so we neglect the negative root of $\rho$. To see the variation of energy density, pressure and energy condition for each of the model the values of constants $B$, $n$, and $\lambda$ are considered as $1$.
\begin{figure}[ht!]
\includegraphics[width=15cm]{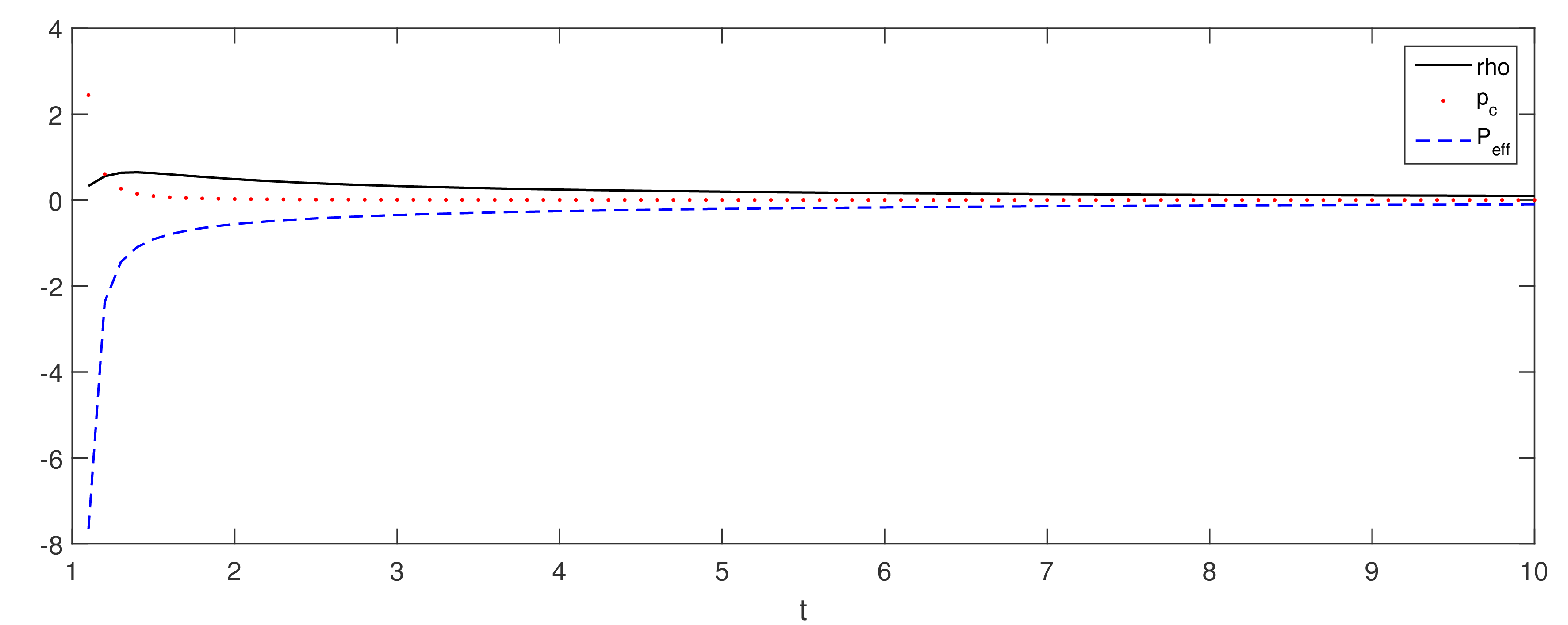}
\caption{Variation of $\rho$, $p^{c}$, and $p_{eff}$ versus $t$ for $B=1$, $\eta= 0.25$, $n=1$ and $\lambda=1$.
\label{fig2}}
\end{figure}
Figure (\ref{fig2}) gives the behaviour of $\rho$, $p_{c}$ and $p_{eff}$ versus cosmic time $t$. $\rho$ is positive throughout the evolution of the universe and is almost equal to zero for large times. At the beginning of the universe, $p_{c}$ is positive and as $t\rightarrow \infty$, $p_{c}\rightarrow 0$. $p_{eff}$ is negative throughout the evolution of the universe and at late times, it has value close to zero. Thus, thermodynamic pressure dominates the creation pressure and leads the expansion of the universe.
\begin{figure}[ht!]
\includegraphics[width=15cm]{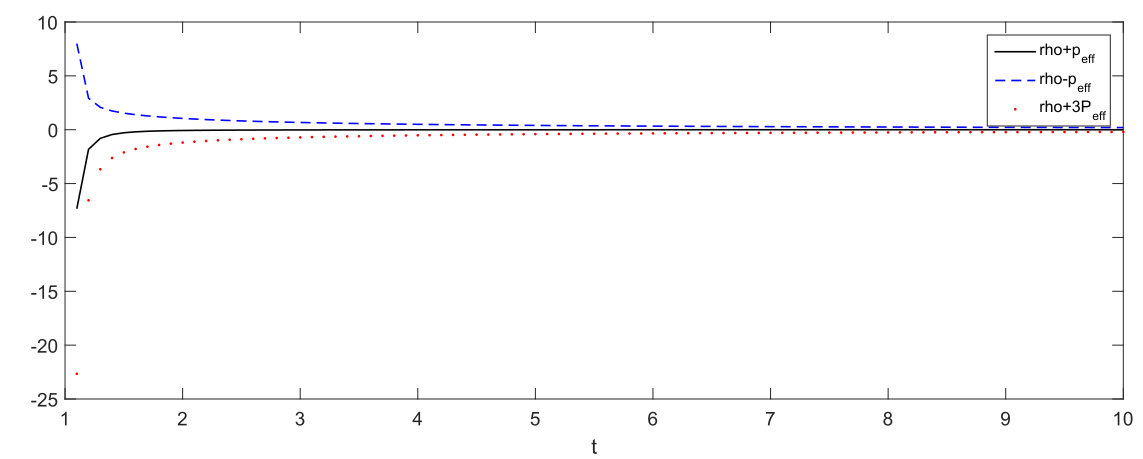}
\caption{Variation of $\rho+p_{eff}$, $\rho-p_{eff}$, and $\rho+3p_{eff}$ versus $t$ for $B=1$, $\eta= 0.25$, $n=1$ and $\lambda=1$.}\label{fig3}
\end{figure}
Figure (\ref{fig3}) depicts the variation of $\rho+p_{eff}$, $\rho-p_{eff}$ and $\rho+3p_{eff}$ with respect to cosmic time $t$. It can be observed that $\rho-p_{eff}>0$ whereas $\rho+3p_{eff}<0$ throughout the evolution of the universe. $\rho+p_{eff}\leq 0$ and as $t\rightarrow \infty$, $\rho+p_{eff}\rightarrow 0$, and $\gamma_{eff}\rightarrow -1$, where $\gamma_{eff}=\frac{p_{eff}}{\rho_{eff}}$ denotes the effective equation of state parameter. And, during the evolution of universe $\gamma_{eff}\leq -\frac{1}{3}$.
\subsection{Model-II}
\begin{figure}[ht!]
\includegraphics[width=15cm]{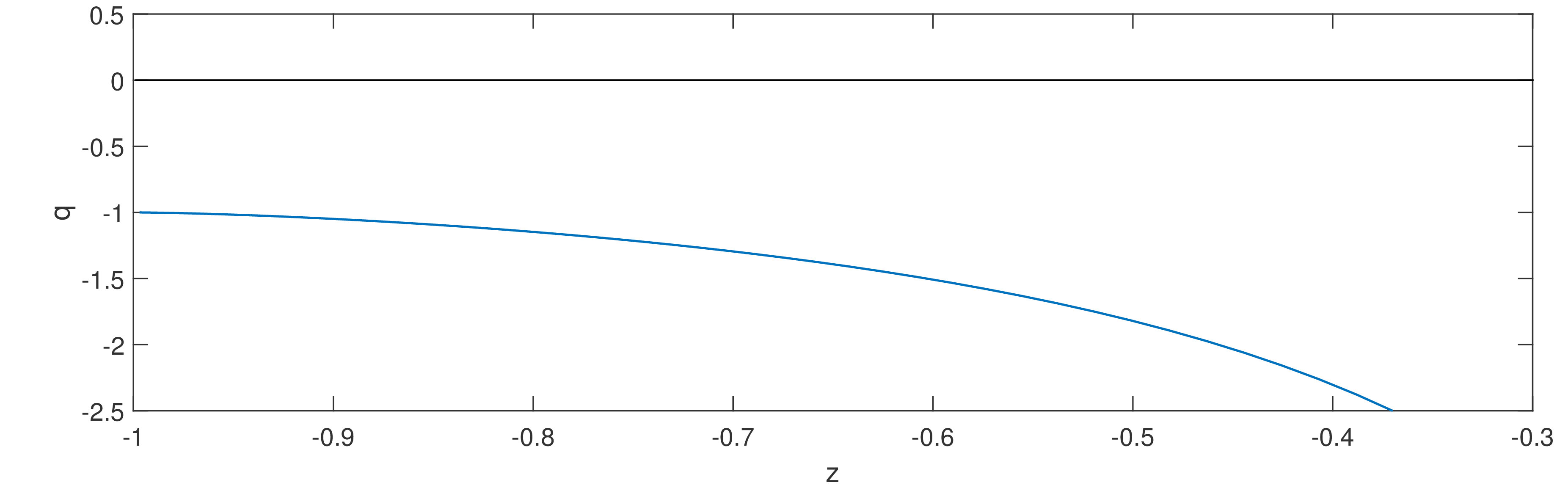}
\caption{Variation of $q$ verses $z$ for $k=1$, $\mu=\frac{\sqrt{3}}{2}$, $\nu=\frac{2}{3}$
\label{fig4}}
\end{figure}

We take the emergent form of scale factor as,
\begin{equation}
a=a_{0}(k+e^{\mu t})^{\nu}
\label{eq24}
\end{equation}
where $a_{0}, k, \mu$ and $\nu$ are any positive constants. For different types of matter in general relativity and in other theories of gravity, this form of scale factor is popular in the cosmological modelling of universe  \cite{70,71,72,73,74,75,76,77}. Using the mechanism of particle creation, a model of emergent universe is formulated in the framework of spatially flat FRW space-time with perfect fluid satisfying barotropic equation of state by considering the universe as a non equilibrium thermodynamical system with dissipation due to particle creation \cite{40}. The emergent universe scenario with modified Chaplygin gas in flat FRW geometrical background has been studied \cite{78}, and it had been concluded that the model does not exhibits emergent scenario at early epochs. The thermodynamical aspects of FRW model of universe with dissipative phenomenon related to effective bulk viscus pressure has been investigated recently \cite{79}. Maity et al \cite{80} have studied the thermodynamic stability of FRW universe having a system of non-interacting diffusive fluids with variable equation of state parameter. It would be interesting to study the behaviour of energy conditions and parameters of expanding universe as par emergent law in the framework of $f(R,T)$ gravity with particle creation and variable chaplygin gas. We get Hubble parameter $H$ as
\begin{equation}
H=\frac{\mu\nu e^{\mu t}}{k+e^{\mu t}}
\label{eq25}
\end{equation}
Deceleration parameter for the emergent universe takes the form,
\begin{equation}
q=-1-\frac{k}{\nu e^{\mu t}}
\label{eq26}
\end{equation}
From equation (\ref{eq26}) it is clear that emergent form of scale factor gives the time dependent $q$ which tends to $-1$ at large times. Figure (\ref{fig4}) gives the variation of $q$ versus $z$. $q$ being always less than $-1$ gives us the super-accelerating universe scenario.

Equation (\ref{eq19}) is a quadratic equation of $\rho$. We get the roots of $\rho$ in terms of $t$ as
\begin{equation}
\rho=-\left[\frac{k\mu^{2}\nu e^{\mu t}}{2(\eta-1)(4\pi+\lambda)(k+e^{\mu t})^{2}}\right]\pm\left[\left(\frac{k\mu^{2}\nu e^{\mu t}}{2(\eta-1)(4\pi+\lambda)(k+e^{\mu t})^{2}}\right)^{2}+\frac{B}{a_{0}^{n}(k+e^{\mu t})^{n-\nu}}\right]^{\frac{1}{2}}
\label{eq27}
\end{equation}

\begin{figure}[ht!]
\includegraphics[width=15cm]{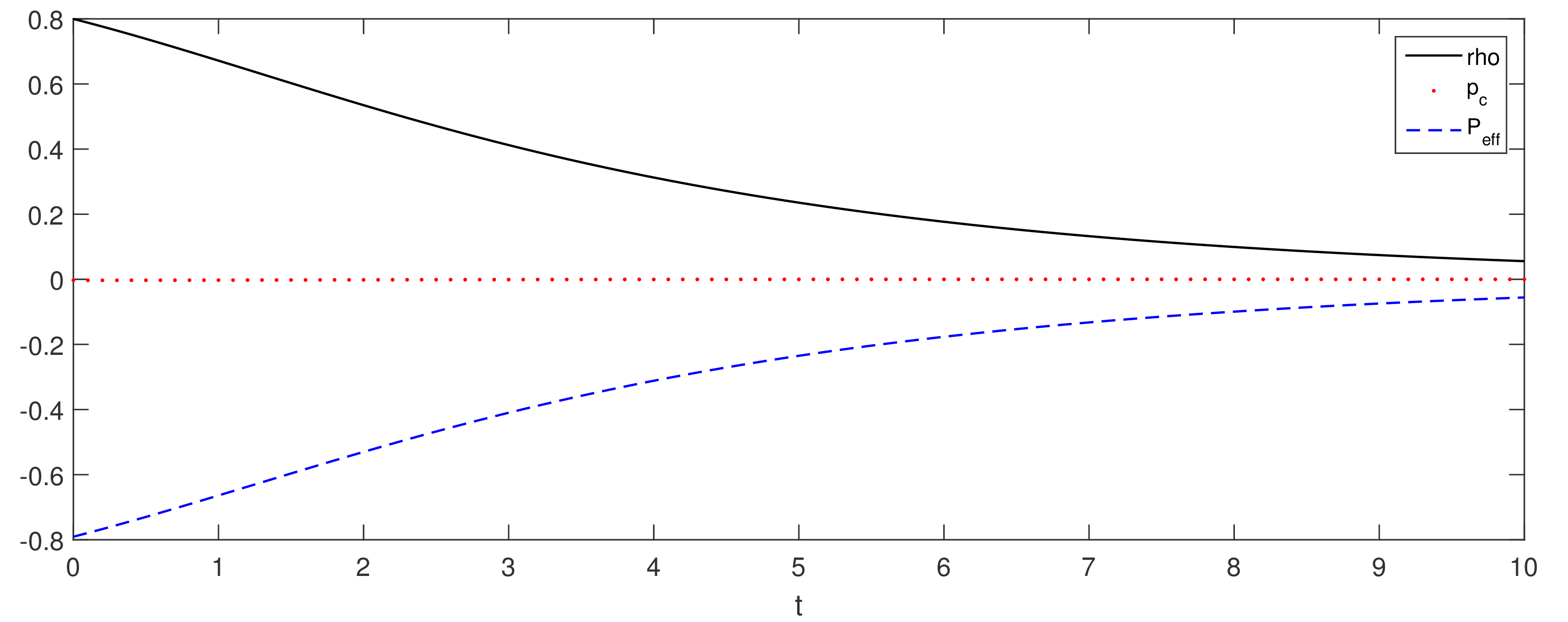}
\caption{Variation of $\rho$, $p^{c}$, and $p_{eff}$ versus $t$ for $B=1$, $\eta= 0.25$, $n=1$, $\lambda=1$, $k=1$, $\mu=\frac{\sqrt{3}}{2}$ and $\nu=\frac{2}{3}$.
\label{fig5}}
\end{figure}
From figure (\ref{fig5}), it can be seen that $\rho$ and $p_{eff}$ has the same value but opposite in sign. From equation (\ref{eq12}), $p_{c}$ is zero. The pressure $p_m$ being negative, causes the expansion of the universe by dominating upon the creation pressure.

\begin{figure}[ht!]
\includegraphics[width=15cm]{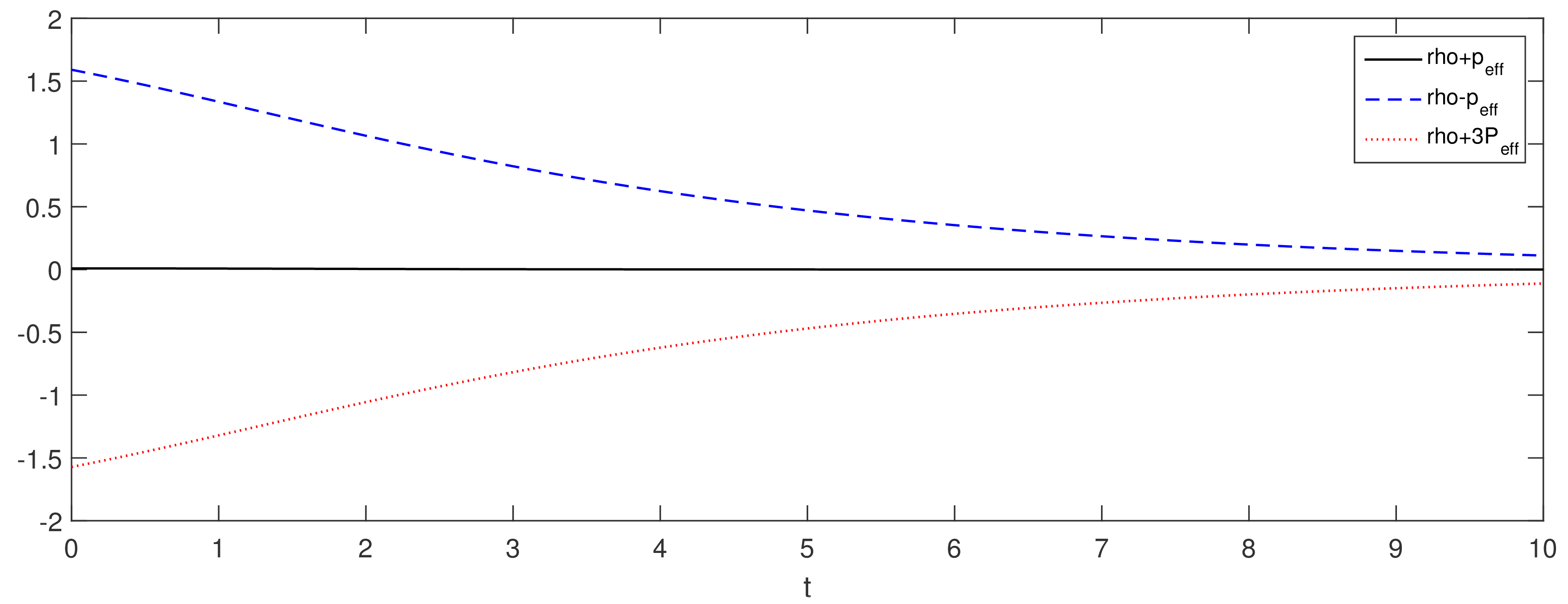}
\caption{Variation of $\rho+p_{eff}$, $\rho-p_{eff}$, and $\rho+3p_{eff}$ versus $t$ for $B=1$, $\eta= 0.25$, $n=1$, $\lambda=1$, $k=1$, $\mu=\frac{\sqrt{3}}{2}$ and $\nu=\frac{2}{3}$.
\label{fig6}}
\end{figure}
One can notice that $\rho$ and $p_{eff}$ have opposite sign, $\rho+p_{eff}$ is always zero, resulting into $\gamma_{eff}= -1$ scenario. On the other hand, from figure (\ref{fig6}), we observe $\rho-p_{eff}\geq 0$ and $\rho+3p_{eff}\leq 0$ throughout the evolution of the universe.
\subsection{Model-III}
\begin{figure}[ht!]
\includegraphics[width=15cm]{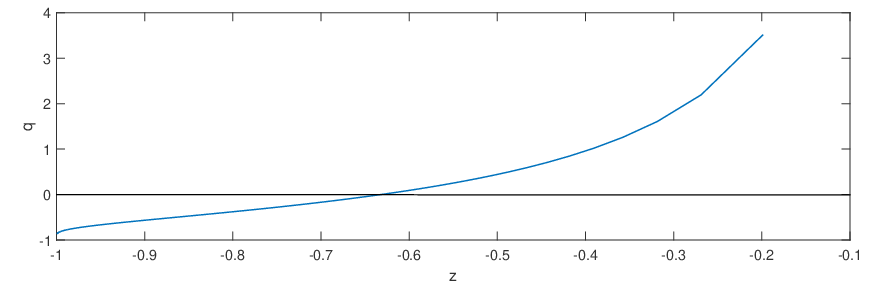}
\caption{Variation of $q$ versus $z$ for $n=0.5$, $m=0.7$
\label{fig7}}
\end{figure}

John D. Barrow have studied the intermediate expansion law in cosmology for the first time in 90s \cite{81,82}. The intermediate form of scale factor gets an exponential function of time as
\begin{equation}
a(t)=exp(mt^{l})
\label{eq28}
\end{equation}
where $m>0$ and $0<l<1$ are constants. With this form of scale factor inflationary universe expands at a rate intermediate between that of the traditional de Sitter inflation and that of the power law inflationary models. Intermediate inflation when $n=\frac{2}{3}$ creates a scale invariant perturbation in Einstein's gravity \cite{81,82}. Measured by observations on CMB, it is found that the intermediate scenario is able to satisfy the bound on scalar spectra index $n_{s}$ and tensor-to-scalar ratio $r$ \cite{83,84}. A lot of works based on intermediate scale factor in isotropic and anisotropic metric backgrounds have been done in different theories of gravity  \cite{82,83,84,85,86,87,88,89,90,91,92,93,94,95,96,97,98}. It will be interesting to study intermediate form of scale factor in the framework of $f(R,T)$ gravity with particle creation with variable Chaplygin gas. \\ For intermediate form of scale factor, we write Hubble parameter $H$ as
\begin{equation}
H=mlt^{l-1}
\label{eq29a}
\end{equation}
Deceleration parameter for the above considered scale factor is given by
\begin{eqnarray}
q=-\frac{\dot{H}+H^{2}}{H^{2}}=-1-\frac{l-1}{mlt^{l}}
\label{eq30}
\end{eqnarray}
Equation (\ref{eq30}) is decreasing function of time and as $t\rightarrow \infty$, $q\rightarrow -1$. Figure (\ref{fig7}) gives the variation of $q$ versus $z$. This form of scale factor gives a time-dependent deceleration parameter which changes it's value from $q>0 $ to $q<0$. In our model, this transition happens at $z_{tr}=-0.6329$.

Equation (\ref{eq19}) is a quadratic equation of $\rho$. We get the roots of $\rho$ in terms of $t$ as

\begin{equation}
\rho=-\left[\frac{ml(l-1)t^{l-2}}{2(\eta-1)(4\pi+\lambda)}\right]\pm\left[\left(\frac{ml(l-1)t^{l-2}}{2(\eta-1)(4\pi+\lambda)}\right)^{2}+\frac{B}{(exp(mt^{l}))^{n}}\right]^{\frac{1}{2}}
\label{eq31}
\end{equation}

\begin{figure}[ht!]
\includegraphics[width=15cm]{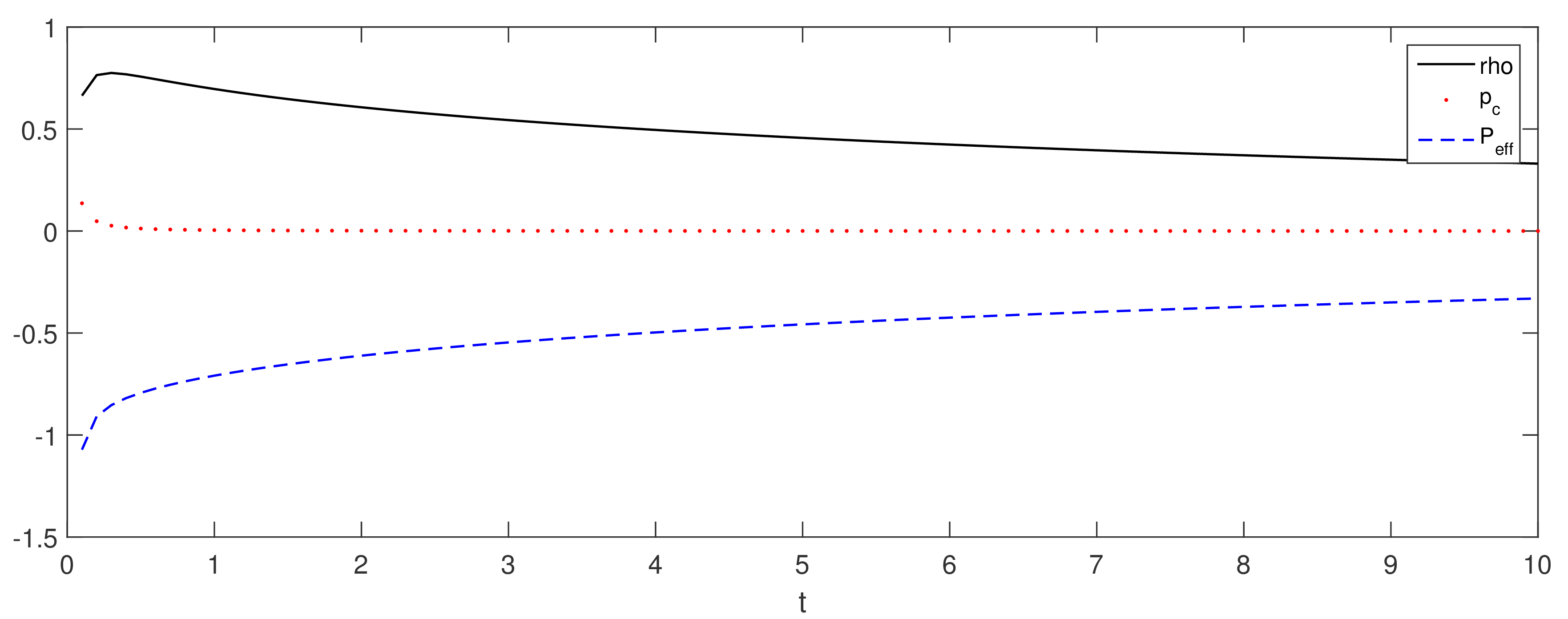}
\caption{Variation of $\rho$, $p^{c}$, and $p_{eff}$ versus $t$ for $B=1$, $\eta= 0.25$, $n=1$, $\lambda=1$, $m=0.7$ and $l=0.5$.
\label{fig8}}
\end{figure}
From figure (\ref{fig8}), we observe that throughout in it's evolution, the universe evolve with positive $\rho$ and $p_{eff}$ is negative. At the beginning of the universe, $p_{c}$ is positive but at late times $p_{c}$ approaches to zero.

\begin{figure}[ht!]
\includegraphics[width=15cm]{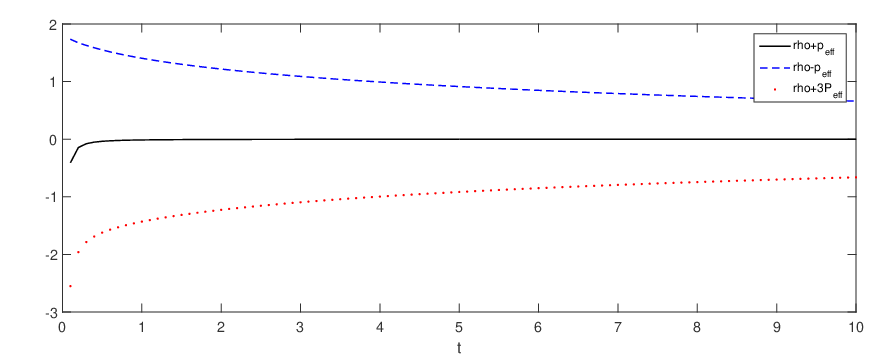}
\caption{Variation of $\rho+p_{eff}$, $\rho-p_{eff}$, and $\rho+3p_{eff}$ versus $t$ for $B=1$, $\eta= 0.25$, $n=1$, $\lambda=1$, $m=0.7$, and $l=0.5$.
\label{fig9}}
\end{figure}

From figure (\ref{fig9}), we observe that $\rho-p_{eff}$ is positive and $\rho+3p_{eff}$ is negative at all times during the evolution of the universe. In the early universe, $\rho+p_{eff}$ is negative but as time evolves $\rho+p_{eff}\rightarrow 0$, yielding $\gamma_{eff}\rightarrow -1$ at late times.

\section{General issues}
The standard point-wise energy conditions are defined as follows \cite{100,101,102}\\
Null energy condition $(NEC) \Leftrightarrow \rho + p\geq 0 $\\
Weak energy condition $(WEC) \Leftrightarrow \rho \geq 0,\ \rho + p\geq 0  $\\
Dominant energy condition $(DEC) \Leftrightarrow \rho\geq 0,\ \rho \pm p\geq 0 $\\
Strong energy condition $(SEC) \Leftrightarrow \rho + 3p\geq 0,\ \rho + p\geq 0 $\\
Violation of NEC will violate all the other energy conditions as well.
\par In model-I, $\rho\geq 0$, $p_{c}\geq 0$ and $p_{eff}<0$ suggest that matter pressure dominates the creation pressure. In the beginning as well as at late times $\rho+p_{eff}<0$. Therefore, model-I violates NEC and hence all the other energy conditions at the beginning of the universe are violated as well; whereas at late times, it satisfies NEC, WEC and DEC but violates the SEC. Source function $(\Gamma)$ (as shown in figure \ref{fig16}) decreases with time but it is positive throughout the evolution of the universe and, at late times, it shows nearly a constant behaviour.
\par In model with emergent form of scale factor $\rho>0$, $p_{eff}<0$ and $p_{c}=0$ throughout the evolution of the universe and, matter pressure dominates the creation pressure. $p_{c}=0$ implies that $\rho=-p_{eff}$. Universe satisfies the NEC, WEC and DEC whereas it violates the SEC throughout its evolution. Source function $(\Gamma)$ shows nearly a constant behaviour for large time $t$.
\par In model with intermediate form of scale factor $\rho>0$, $p_{eff}<0$ throughout the evolution of the universe. At the beginning of the universe $p_{c}$ is positive but as universe evolves creation pressure becomes zero which results into $\rho=-p_{eff}$ at late times. Also, matter pressure dominates the creation pressure. The model shows transition of deceleration parameter from decelerated phase to accelerated phase at $z_{tr}=-0.6329$. Source function $(\Gamma)$ is positive throughout the evolution of the universe and at late times, it shows nearly a constant behaviour with very small value.\\
The emergent form of scale factor in higher derivative theory in flat FLRW geometrical background shows no particle production and satisfies the SEC \cite{49}.\, whereas the emergent scenario case in the present $f(R,T)$ theory with particle creation mechanism have particle production with constant rate and the universe is violating SEC also.\\
State-finder diagnostic pair is a diagnos \cite{99}, which allows us to differentiate between various dark energy models. Through the higher derivatives of $a$, $H$ and $q$, the state-finder diagnostic pair $\{r,s\}$ being the geometrical parameters in the sense that it is constructed directly from the space-time
metric, allows us to investigate the expansion dynamics of the universe. $\{r,s\}$ parameters  are defined as
\begin{equation}
r=\frac{\dddot{a}}{aH^{3}}  and s=\frac{r-1}{3(q-0.5)}
\label{eq29}
\end{equation}
The orientation lines in the $s-r$ plane having $\{0,1\}$ as a fixed point, highlights the spatially flat FRW universe standard $\Lambda$CDM model whereas $\{1,1\}$ is a fixed point in a standard $SCDM$ model. The position of the fixed point $\{s,r\}$ has been calculated in order to calculate the diverging or converging behaviour of our dark energy model with respect to the $SCDM$ or $\Lambda CDM$ model.\\
The $\{r,s\}$ parameters for model-I are obtained as,
\begin{equation}
r=\frac{6(t^{3}-1)^{2}}{(1+2t^{3})^{3}}
\label{eq33}
\end{equation}
\begin{equation}
s=\frac{2(8t^{9}+6t^{6}+18t^{3}-5)}{16t^{9}+6t^{3}+5}
\label{eq34}
\end{equation}

\begin{figure}[ht!]
\includegraphics[width=15cm]{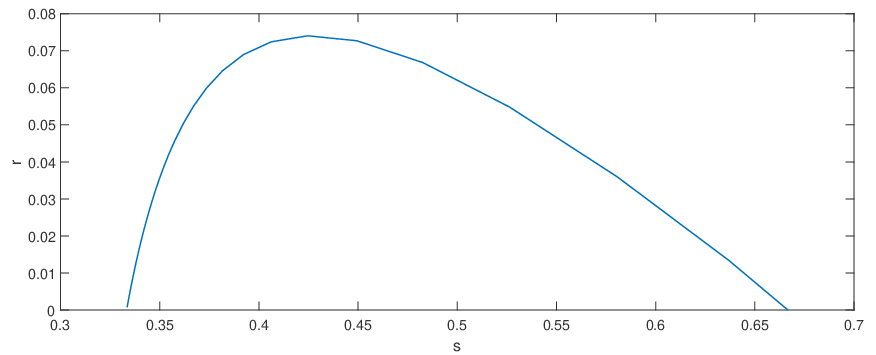}
\caption{Variation of $r$ versus $s$.
\label{fig10}}
\end{figure}

\begin{figure}[ht!]
\includegraphics[width=15cm]{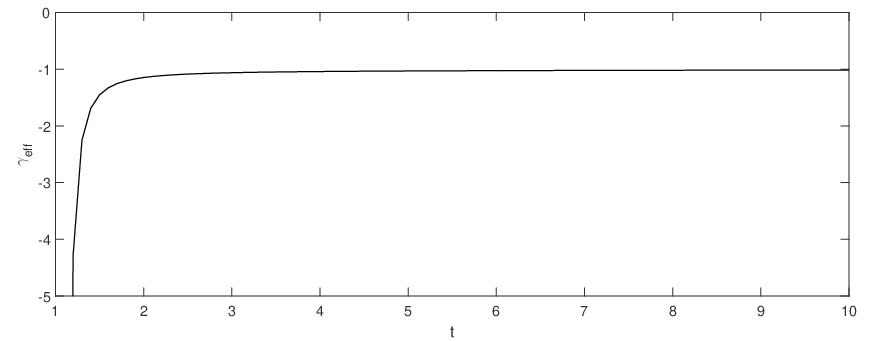}
\caption{Variation of $\gamma_{eff}$ versus $t$ for $B=1$, $\eta=0.25$, $n=1$ and $\lambda=1$.
\label{fig11}}
\end{figure}
\begin{figure}[ht!]
\includegraphics[width=15cm]{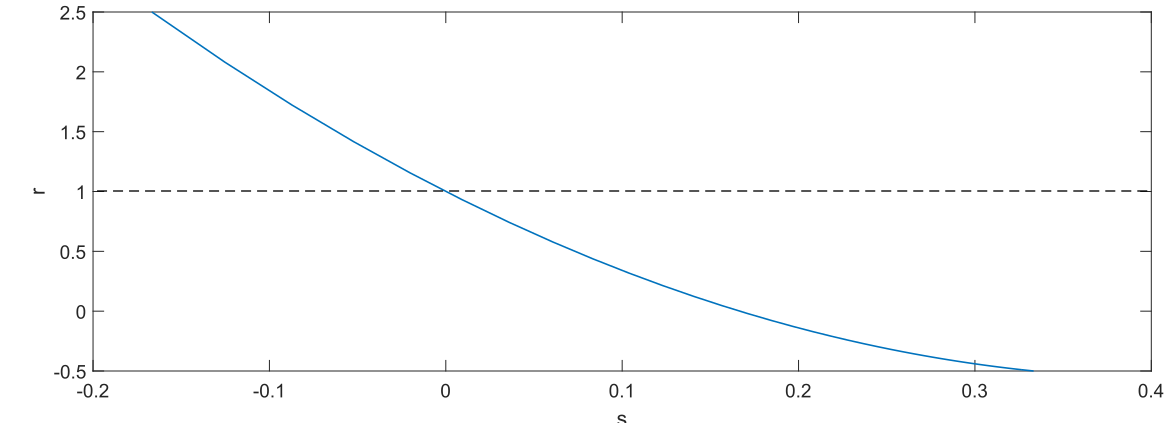}
\caption{Variation of $r$ vs $s$ for $k=1$, $\mu=\frac{\sqrt{3}}{2}$ \& $\nu=\frac{2}{3}$.
\label{fig12}}
\end{figure}
\begin{figure}[ht!]
\includegraphics[width=15cm]{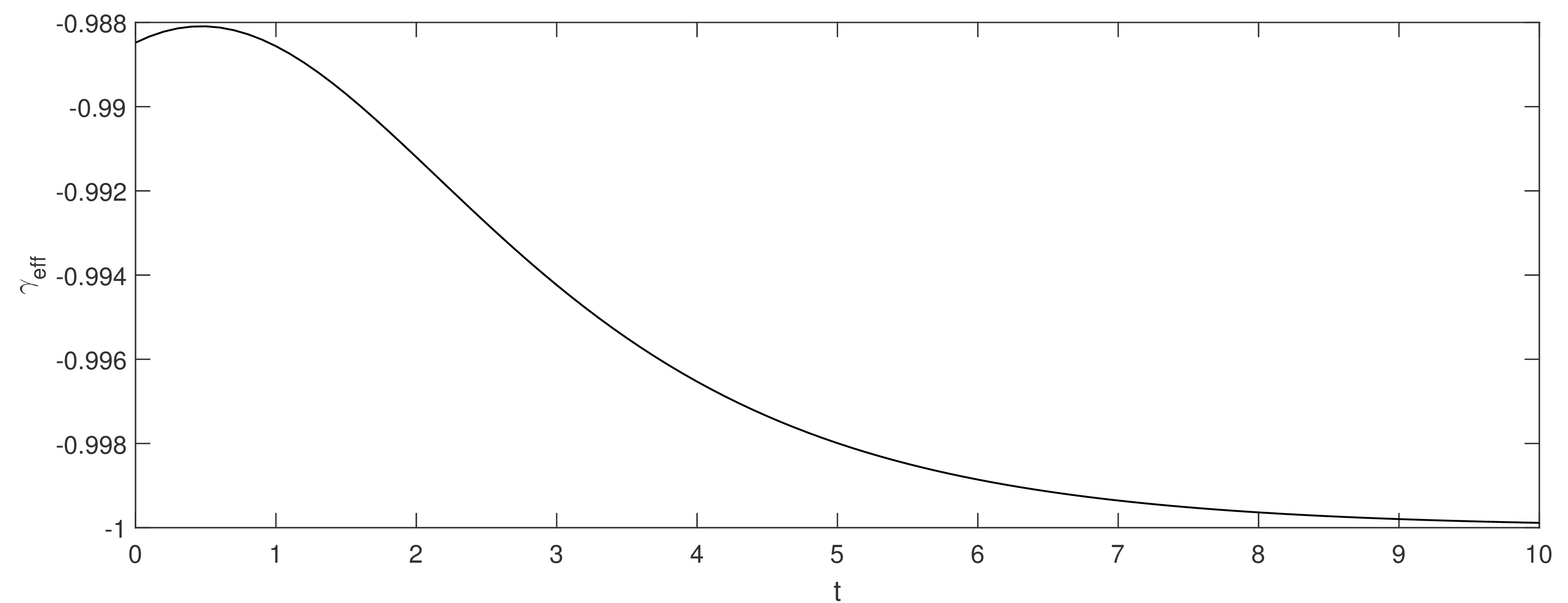}
\caption{Variation of $\gamma_{eff}$ vs $t$ for $B=1$, $\eta=0.25$, $n=1$ and $\lambda=1$, $k=1$, $\mu=\frac{\sqrt{3}}{2}$ \& $\nu=\frac{2}{3}$.
\label{fig13}}
\end{figure}

From figure (\ref{fig10}), it can be seen that the trajectory in $\{r,s\}$ does not approach to find fixed point of $\Lambda CDM$ model. From figure (\ref{fig11}), we observe that $\gamma_{eff}\rightarrow -1$ at late times, giving us the late time accelerating scenario of the universe.\\
The $\{r,s\}$ parameters for model-II are obtained as follows

\begin{eqnarray}
r=\frac{\dddot{a}}{aH^3}=\frac{(k+\nu e^{ut})[k+e^{ut}(\nu-1)]}{\nu^2 e^{2ut}}
\label{eq35}
\end{eqnarray}
\begin{eqnarray}
s=\frac{r-1}{3(q-0.5)}=\frac{-2k(k+2\nu e^{ut})+2e^{ut}(k+\nu e^{ut})}{3\nu e^{ut}(2k+3\nu e^{ut})}
\label{eq36}
\end{eqnarray}
From figure (\ref{fig12}), we observe that for the suitable choices of parameters, the trajectories in $\{r,s\}$ plane passes through the fixed point $\{1,0\}$ of $\Lambda$CDM model. From figure (\ref{fig13}), we gets $\gamma_{eff}\rightarrow -1$ at late times, giving us the late time accelerating scenario of the universe.

\begin{figure}[ht!]
\includegraphics[width=15cm]{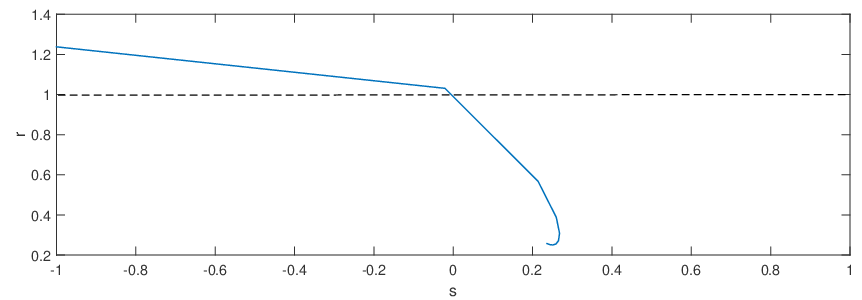}
\caption{Variation of $r$ vs $s$ for $m=0.7$ \& $l=0.5$.
\label{fig14}}
\end{figure}

\begin{figure}[ht!]
\includegraphics[width=15cm]{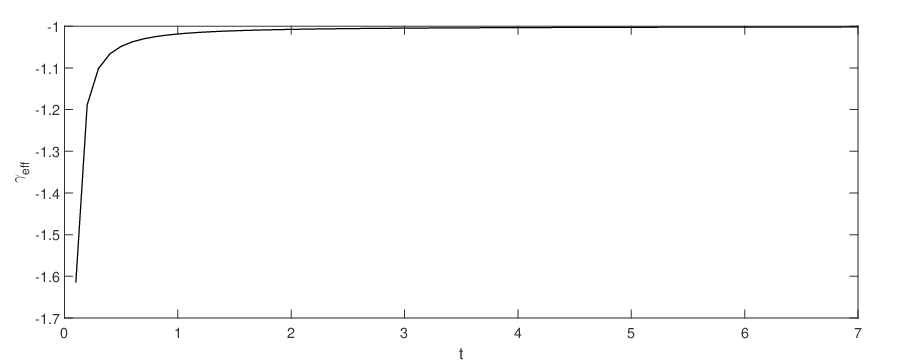}
\caption{Variation of $\gamma_{eff}$ vs $t$ for $B=1$, $\eta=0.25$, $n=1$ and $\lambda=1$, $m=0.7$ \& $l=0.5$.
\label{fig15}}
\end{figure}

\begin{figure}[ht!]
\includegraphics[width=15cm]{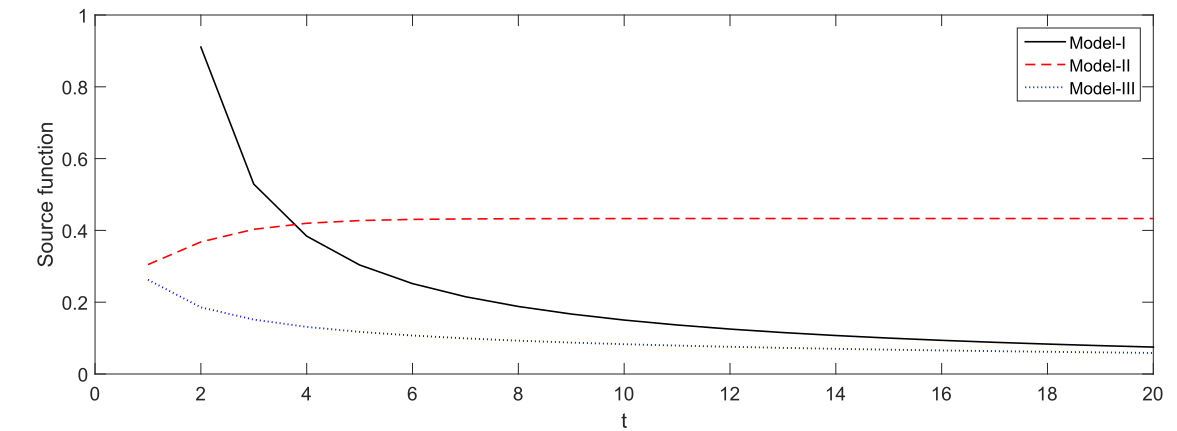}
\caption{Variation of source function $(\Gamma)$ versus $t$
\label{fig16}}
\end{figure}

The $\{r,s\}$ parameters for model-III are as follows,

\begin{eqnarray}
r=\frac{\dddot{a}}{aH^{3}}=\frac{(l-1)[(l-2)+2mlt^{l}]+mlt^{l}[(l-1)+mlt^{2}]}{(ml)^2t^{2l}}
\label{eq37}
\end{eqnarray}

\begin{eqnarray}
s=\frac{r-1}{3(q-0.5)}=\frac{2(l-1)[(l-2)+3mlt^{l}]}{3ml[6(1-l)-3mlt^{l}]}
\label{eq38}
\end{eqnarray}
From figure (\ref{fig14}), we observe that for the suitable choices of parameters, the trajectories in $\{r,s\}$ plane passes through the fixed point $\{1,0\}$ of $\Lambda$CDM model for the intermediate expansion law model.  From figure (\ref{fig15}), we gets $\gamma_{eff}\rightarrow -1$ at late times in the universe with intermediate from of scale factor.

Figure (\ref{fig16}) shows the behaviour of source function $\Gamma$ with respect to cosmic time $t$. Interestingly, in all the three models proposed here, universe is having particle production scenario having almost constant value at late times.

\section{Conclusions}
We have considered three different anstaz of scale factors to study the particle creation mechanism in the flat FLRW geometrical background with variable Chaplygin gas form of equation of state parameter in $f(R,T)$ gravity. A negative pressure fluid is the mechanism responsible for accelerated phase of universe expansion in standard cosmology. Particle creation mechanism in the considered $f(R,T)$ gravity for variable Chaplygin gas allows us to understand the particle production, annihilation for the considered form of scale factor for the expanding model of the universe.\\
The universe model with scale factor of form $a(t)=t^2-\frac{1}{t}$ is an ever accelerating universe whereas emergent universe (Model-II) expand with super-acceleration. Universe with intermediate form of scale factor does transition from decelerating phase to accelerating phase. \\
In all the three considered cases of scale factor governed universe, matter pressure dominates the creation pressure to give negative effective pressure $p_{eff}$ and hence becomes the cause for the expansion of the universe. These models of universe satisfies  NEC, WEC and DEC whereas violates the SEC during its evolution. Therefore, considered models in this paper having $\gamma_{eff}<-\frac{1}{3}$ are capable of explaining the current phase of expansion of the universe. \\
Positive $\Gamma$ indicates particle production, negative $\Gamma$ suggest particle annihilation and vanishing $\Gamma$ shows no particle production. The cosmological models of universe have $\Gamma>0$, during their evolution. So, all the considered form of scale factors (in this paper) in $f(R,T)$ gravity with variable Chaplygin gas in the flat FLRW geometrical background shows particle production with almost constant behaviour at late times with very small value. \\
Solutions and results demonstrated in this paper may be useful for better understanding of the characteristics of particle creation during the evolution of the universe in the $f(R,T)$ gravity in flat FLRW geometrical background with perfect fluid, and this work may be further extended with other forms of equation of state parameters as well as anisotropic geometrical backgrounds with care of, as the universe is presently expanding with acceleration and having a history of signature flipping in terms of deceleration parameter $q$ from positive to negative, as per recent cosmological observations \cite{1,2,3,4,103}.

\end{document}